\documentclass{article}

\usepackage{amsthm,amsmath}
\usepackage[utf8]{inputenc} 

\def\includegraphics{}

\usepackage{graphicx}
\usepackage[T1]{fontenc}
\usepackage[utf8]{inputenc}
\usepackage{authblk}
\begin{document}



\title{Analysis of Triplet Motifs in Biological Signed Oriented Graphs Suggests a Relationship Between Fine Topology and Function}

\author[1]{Alberto Calderone*}
\author[1]{Gianni Cesareni}
\affil[1]{
Bioinformatics and Computational Biology Unit\\
Department of Biology\\
University of Rome 'Tor Vergata'\\
Via della Ricerca Scientifica, 1 - 00133 - Rome - Italy
\\
contact: sinnefa@gmail.com}

\maketitle

\begin{abstract} 
Background:
Networks in different domains are characterized by similar global characteristics while differing in local structures. To further extend this concept, we investigated network regularities on a fine scale in order to examine the functional impact of recurring motifs in signed oriented biological networks. In this work we generalize to signaling networks some considerations made on feedback and feed forward loops and extend them by adding a close scrutiny of \textit{Linear Triplets}, which have not yet been investigate in detail.

Results:
We studied the role of triplets, either open or closed (Loops or linear events) by enumerating them in different biological signaling networks and by comparing their significance profiles. We compared different data sources and investigated the fine topology of protein networks representing causal relationships based on transcriptional control, phosphorylation, ubiquitination and binding. Not only were we able to generalize findings that have already been reported but we also highlighted a connection between relative motif abundance and node function. Furthermore, by analyzing for the first time \textit{Linear Triplets}, we highlighted the relative importance of nodes sitting in specific positions in closed signaling triplets. Finally, we tried to apply machine learning to show that a combination of motifs features can be used to derive node function.

Availability:
The triplets counter used for this work is available as a Cytoscape App and as a standalone command line Java application.
\\http://apps.cytoscape.org/apps/counttriplets
\\
\textbf{Keywords:} Graph theory, graph analysis, graph topology, machine learning, cytoscape

\end{abstract}


%



\section*{Background}
Biological networks share global characteristics such as a relatively short path between any two nodes (small-world) and a node degree distribution which follows a power-law \cite{Girvan2002}. The recurrence of these statistical features can be used to assess network similarity on a global scale. On the other hand, while natural networks in general tend to have similar global characteristics, they differ in local structures \cite{Maslov2002}. This characteristic can be used to compare network in general and biological processes in physiology and pathology \cite{Calderone2016}.

In computational network biology, other than assessing similarities one can investigate the possible relationships between topology and molecular function. The first and simplest approach is the analysis of nodes neighbors \cite{Schwikowski2000}. Other approaches are based on the premise that functional modules are assemblies of cellular elements linked to a common biological function \cite{Hartwell}. In this case, functions are not associated to single genes but are derived from groups of genes. Some algorithms can detect molecular complexes \cite{Bader2003}, while others can handle larger, albeit physically looser, functional structures such as signaling pathways \cite{Scott2006}.

From a more granular perspective, one can inspect network fine structure by analyzing the topology of smaller groups of interconnected nodes (terns, quartets, etc...) that frequently recur (network motifs) in biological networks \cite{Vazquez2004}. In general, motifs that are more frequently observed than expected by chance are deemed to underlie relevant properties.

From a biological perspective, it was proposed that different network motifs underlie specific functions in gene expression where they can, for instance, modulate the expression kinetics of genes responding to signals propagating from the membrane to the nucleus \cite{Shen-Orr2002}\cite{Milo2002}. Among these motifs, triangles were studied and characterized from a functional perspective in the context of transcriptional networks. For example, feedback loops play a self-regulatory role in the $\lambda$-phage lysogenic cycle \cite{McAdams1995} while feed forward loops can modulate the speed and timing of gene expression in general \cite{Shen-Orr2002}\cite{Mangan2003}. Due to their important roles, feed forward loops are particularly frequent in gene regulatory networks and more frequent than feedback loops \cite{Vinayagam2014}. It is not clear whether these regularities can be generalized to a wider spectrum of biological networks such as, for instance, signaling networks.

To assess the functional relevance of local properties of signaling networks, we investigated the importance of recurring motifs in signed oriented biological networks a kind of analysis which has been partially hampered by the lack of suitable curated data.

Well established interaction databases such as the one curated by the MIntAct project \cite{Orchard2014} and mentha \cite{Calderone2013} capture and store information on physical protein-protein interactions. However, these resources do not yet annotate causal relationships which are essential to capture the information flow in signaling networks. To this end, we extracted data from the SIGNOR database \cite{Perfetto2015} and compared it against other resources annotating causal relationships such as KEGG \cite{Kanehisa2000} and SignaLink \cite{Fazekas2013}. In addition, we also considered a manually curated flat file compiled by the group of Edwin Wang \cite{Zaman2013}. SIGNOR was also used to perform specific analyses requiring annotation on the interaction type: transcriptional, phosphorylation, ubiquitination and binding.
 
In order to investigate whether network motifs are related to node function we applied machine learning to predict molecular function of a specific node from a combination of the abundance of each network motif. This approach suggested a relationship between fine topology and function.

The novelty of our study resides in the analysis of causal interaction  data extracted from four different resources annotating causal relationships. By this approach we could extend the observations on transcriptional regulatory networks \cite{Mangan2003}\cite{Shen-Orr2002}\cite{Milo2002} to signaling networks in general. In addition to confirming and strengthening, on a larger scale, previously reported findings we eventually formulate more general conclusions. Our study not only compares networks from different resources, but it also considers different kinds of interactions (graph edges): transcriptional regulation, phosphorylation, ubiquitination and binding. These detailed analysis allowed us to conclude that certain protein classes, such as receptors and phosphatases are preferentially associated to specific network motifs. Furthermore, we investigate for the first time the role of \textit{Linear Triplets} which give information on the role played by a node sitting in a specific place inside a triangle.

In order to promote these kind of analyses for other higher coverage networks that might become available in the future, we release standalone command line tool which can also work as a a Cytoscape App (http://apps.cytoscape.org/apps/counttriplets) (Supplementary Material).

\section*{Methods}
All the analyses started from an exhaustive enumeration of network motifs. To this end, we developed a piece of software in Java using the JUNG library \cite{OMadadhain2005}. We packed our software in a .jar file, which can be either run as a standalone tool or installed in Cytoscape.

Using our application we counted motifs consisting of three elements which we called Triplets in order to distinguish them from triads, which is the de-facto name for motifs in oriented, but not signed, networks. In particular, we counted \textit{Closed Triplets} (triangles) and \textit{Linear Triplets} (open triangles, three nodes in line).

The number of motifs in a complete signed oriented graph is given by the following formula:

\begin{eqnarray}
\label{eq:tripletscount}
\binom{n}{3}*(l^3-(1+(3*(l-1)))-d*k))
\end{eqnarray}

\begin{itemize}
\item n is the number of nodes in a complete signed oriented graph
\item k is the number of colors an edge can have (red and blue, activation and inhibition)
\item d are the possible states of an edge (from A to B, from B to A, absent)
\item l is d+k. It is the number of possible labels an edge can have so it is the k colors plus the possible effects d: present right-to-left activation, present left-to-right activation, present right-to-left inhibition, present left-to-right inhibition, absent).
\end{itemize}

$\binom{n}{3}$ are all possible triangles. $l^3$ is all the possible configuration three edges can have. From these we need to remove the empty triangle, the 1 in the formula, all the possible configurations with only one edge (3*(l-1)) and all the isomorph triangles (d*k).

The table (Table~\ref{tablecount}) shows how the total number of \textit{Closed Triplets} (triangles) and \textit{Linear Triplets} (open triangles, three nodes in line) grows with the number of nodes considered. 

\begin{table}[!t]
\centering
\label{tablecount}
\begin{tabular}{|c|c|}
\hline
Nodes & Triplets\\\hline
3 & 106\\
4 & 424\\
5 & 1060\\
6 & 2120\\
7 & 3710\\
8 & 5936\\
9 & 8904\\
10 & 12720\\\hline
\end{tabular}
\caption{Total number of triplets found in a complete oritented signed graphs. This table lists how many triplets can be counted in a complete signed oriented graph calculated with Eq~\ref{eq:tripletscount}}
\end{table}

From the table we can see the exponential growth of  the possible configurations. Luckily, the analyzed networks are not complete and such enumeration can be performed exhaustively without computational problems. We used this formula to check the correctness of the application we used in our analysis.

\textit{Linear Triplets} can give detailed information on the role of each node in a Closed Triplet as they represent a way to only look at the ingoing/outgoing edges of a node. Put simply, this second motif class is a somewhat finer measurement of \textit{Closed Triplets}.

\subsection*{Abundance and Significance Analysis}

For our preliminary analyses we looked at motif abundance by plotting motif frequency histograms and thus making motifs abundance comparable through datasets.
In order to visualize and compare network motifs profiles we adopted the same strategy used in previous studies \cite{Milo2004} \cite{Wong2012}. z-scores were normalized as shown in the following formulae:

\begin{eqnarray}
\label{eq:z}
Z_{i}=\frac{N_{real}-mean(N_{random})}{std.dev.(N_{random})}\\
\label{eq:sp}
SP_{i}=\frac{Z_{i}}{(\sum_{j=1}^{M}Z_{j}^2)^{\frac{1}{2}}}
\end{eqnarray}

Where $N_{real}$ is the number of occurrences of a given motif in the real network, $N_{rand}$ is the average number of occurrences of a given motif in randomly generated networks (5,000 in this analysis) created by preserving in and out degrees and edge signs ratio. $M$ is the number of counted motifs.
The $SP$ (significance profile) highlights the relative significance of a motif rather than its absolute significance \cite{Wong2012}, allowing for comparison of networks of different sizes (Table~\ref{tablecount}). Motifs in large networks will otherwise have higher z-scores than in small networks.

\subsection*{Compared Data Sources} 
The four networks analyzed were processed as follows:
\begin{enumerate}
\item SIGNOR \cite{Perfetto2015}: archives direct and indirect causal interactions between different kinds of nodes. We only considered direct interactions between proteins.
\item KEGG \cite{Kanehisa2000}: contains metabolic, signaling and other kinds of pathways. We parsed pathways containing the word "signaling" in their names in order to extract directed activations and inhibitions interactions. 
\item SignaLink \cite{Fazekas2013}: stores direct and indirect causal interactions between proteins and RNAs. We selected only direct interactions between proteins where the effect is different from "unknown".
\item Edwin Wang network \cite{Zaman2013}: annotates positive, negative and physical interactions between genes. We only considered "pos" ad "neg" interactions, excluding interactions only reported as physical.
\end{enumerate}

Other than analyzing different data sources, we extended our analysis to four subnetworks extracted from the SIGNOR database. We derived a network with transcription interactions, one with (de)phosphorylation interactions, one with ubiquitination interactions and one with binding interactions.

\begin{table}[!t]
\centering
\begin{tabular}{|l|r|r|r|r|r|}
\hline
  & {\bf Nodes} & {\bf Edges} & {\bf Activation Ratio} & {\bf Both Signs** Ratio} & {\bf Transitivity}\\\hline
SIGNOR & 2949 & 6666 & 0.627 & 0.015 & 0.064\\
SignaLink & 752 & 1602 & 0.976 & 0.001 & 0.109\\
KEGG & 693 & 1226 & 0.784 & 0.009 & 0.068\\
Edwin Wang & 6005 & 41052 & 0.807 & 0.000 & 0.124\\
Transcription* & 632 & 855 & 0.726 & 0.001 & 0.031\\
(de)Phosphorylation* & 1597 & 3864 & 0.555 & 0.026 & 0.064\\
Ubiquitination* & 197 & 199 & 0.236 & 0.005 & 0.036\\
Binding* & 1840 & 2437 & 0.749 & 0.012 & 0.050\\\hline
\end{tabular}
\label{table1}
\caption{Signaling Networks used in this work. All four networks have similar activation ratio, about 80\%. This homogeneity is not preserved in SIGNOR subnetworks. The phosphorylation subnetwork activations ratio is only 55\%, while in the ubiquitination subnetwork 76\% of interactions are inhibitions. ** interactions with one direction that has both effects on the target node at the same time.
* subnetworks derived filtering the SIGNOR global network.}
\end{table}

\subsection*{Combining Features to Infer Molecular Functions}
We used a supervised machine learning approach to assess the feasibility of classifying proteins according to their motifs abundance profile. In particular, we used the caret package \cite{Kuhn2008} to perform various analysis.

We used Random Forest in order to inspect the relative importance of one motif over the others in determining a node function. We also had to take into account the fact that the collected data is very sparse and unbalanced, i.e. most of the nodes occur in only few motifs, while others have some motifs that appear more often than others by more than one order of magnitude. These two issues are the simple consequence of the different emphasis given in data curation.

We addressed sparseness and unbalance by predicting missing values with multiple linear models where each feature is predicted in function of the other columns. We created these linear models with features selected through a 10-fold cross validation applying a stepwise Akaike information criterion \cite{Akaike1998} to derive the best combination of variables. On average we obtained a $R^2$ of 0.65.

\subsection*{Motifs Nomenclature}
\begin{figure}[htp]
\centering
\includegraphics[scale=0.20]{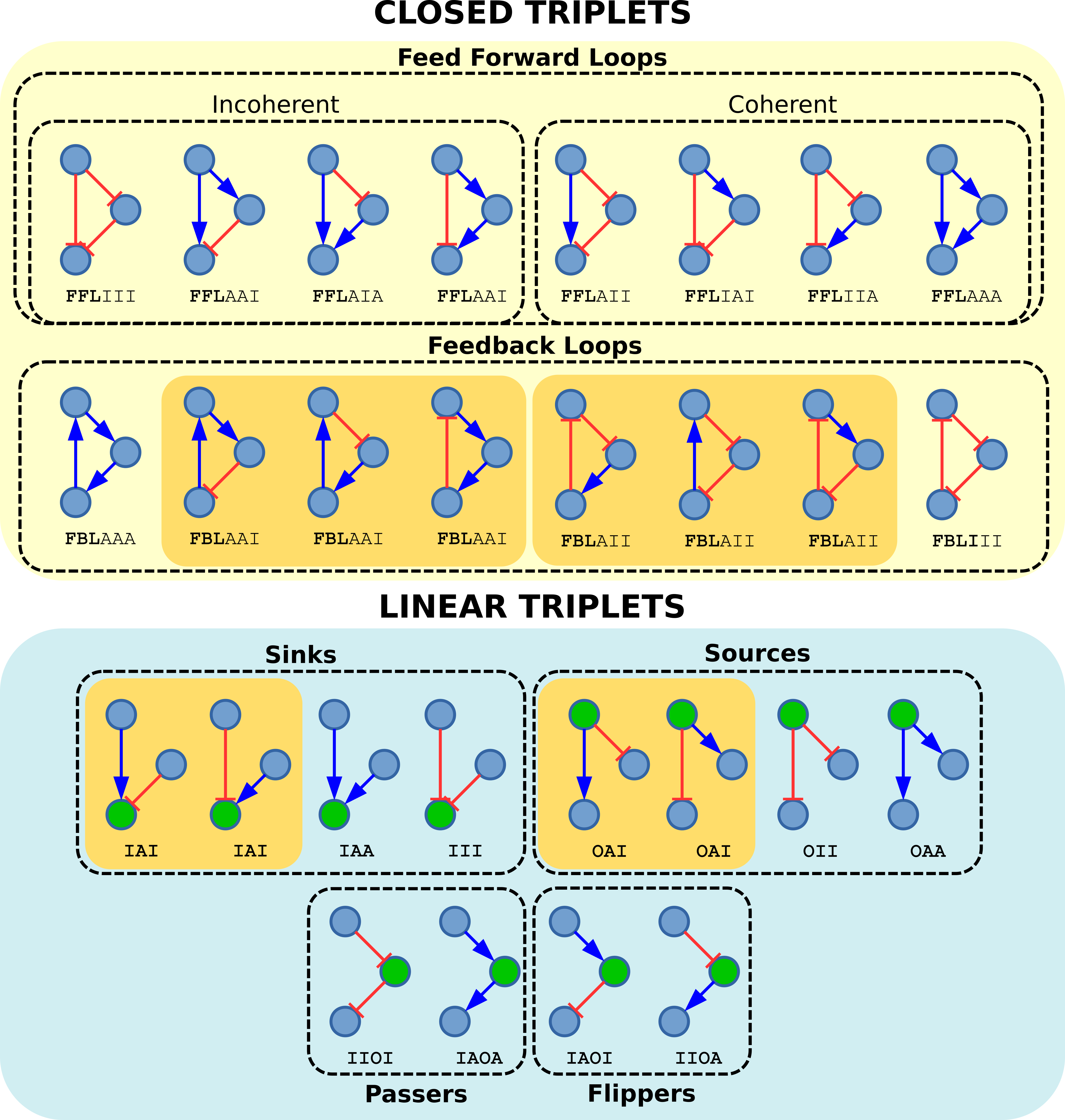}
\caption{{\bf Classification of triplet network motifs} The two main classes are colored in yellow (\textit{Closed Triplets}) and blue (\textit{Linear Triplets}). Motifs highlighted in orange are isomorphism and thus indistinguishable. Incoherent loops are loops where the target node receives two discordant signals while coherent loops are those where the target node receives two concordant signals. \textit{Linear Triplets} are grouped into 4 classes, named according to the incoming and outgoing signals experienced by the central node. Sinks and Sources receive or emit two signals respectively, Passers echo the received signals while Flippers invert the input sign.}
\label{Figure1}
\end{figure}

In defining each motif we need to consider edge directions and signs. The nomenclature used for \textit{Closed Triplets} is based on the number of activations and inhibitions contained in a motif. Labels assigned to feedback loops consist of FBL followed by a number of A's and I's equal to the number of activations (A) and inhibitions (I). This class of motifs contains many isomorphisms, as they are rotations of the same configurations. For example, FBLAAI is indistinguishable from all the other motifs highlighted in the orange area in Fig~\ref{Figure1}.

Differently, for feed forward loops, where it is clear which node is the source (two outgoing edges) and which node is the target (two ingoing edges) in the triangle, we used the label FFL (feed forward loop) followed by an ordered sequence of three letters representing the three effects in the triangle: XYZ where X is the effect from source node to target node, Y in the effect from source node to the intermediate node (one ingoing and one outgoing edge) and Z is the effect from the intermediate node to target node.

For \textit{Linear Triplets} we labeled each configuration taking as a reference the central node (green node Fig~\ref{Figure1}) and describing the two incident edges. These labels can contain three or four characters. We used this convention for \textit{Linear Triplets} so that the nature of the central node is preserved: if the label has three characters, than the motif is a Sink or a Source (with two ingoing or two outgoing edges, Fig~\ref{Figure1}), if it has foud characters it is a Passer, where the output effect is identical to the input effect, or a Flipper, if the output effect changes. For example, OII means that the central node has two outgoing (O) inhibitions (I) while IIOA means that the central node has an ingoing (I) inhibition (I) and an outgoing (O) activation (A). 

\section*{Results}
Our analysis relied on causal information extracted from three online repositories: SIGNOR \cite{Perfetto2015}, KEGG \cite{Kanehisa2000} and SignaLink \cite{Fazekas2013} and a manually curated network by the group of Edwin Wang \cite{Zaman2013}. As shown in (Table~\ref{table1}) these four network differ in node and edges numbers but have similar ratio of activation and inactivation edges. This first comparison implies that, no matter the specific compilation of signaling networks in different curation efforts, approximately 80\% of interactions are activations. It is interesting to notice that such homogeneity among data sources is not preserved in subnetworks derived from SIGNOR. In the phosphorylation subnetwork, the activation ratio is only slightly in favor of activations, 55\%, while in the ubiquitination subnetwork 76\% of interactions are inhibitions. This variation can be seen as a first sign that functionally different networks have different structures.

\begin{figure}[htp]
\centering
\includegraphics[scale=0.28]{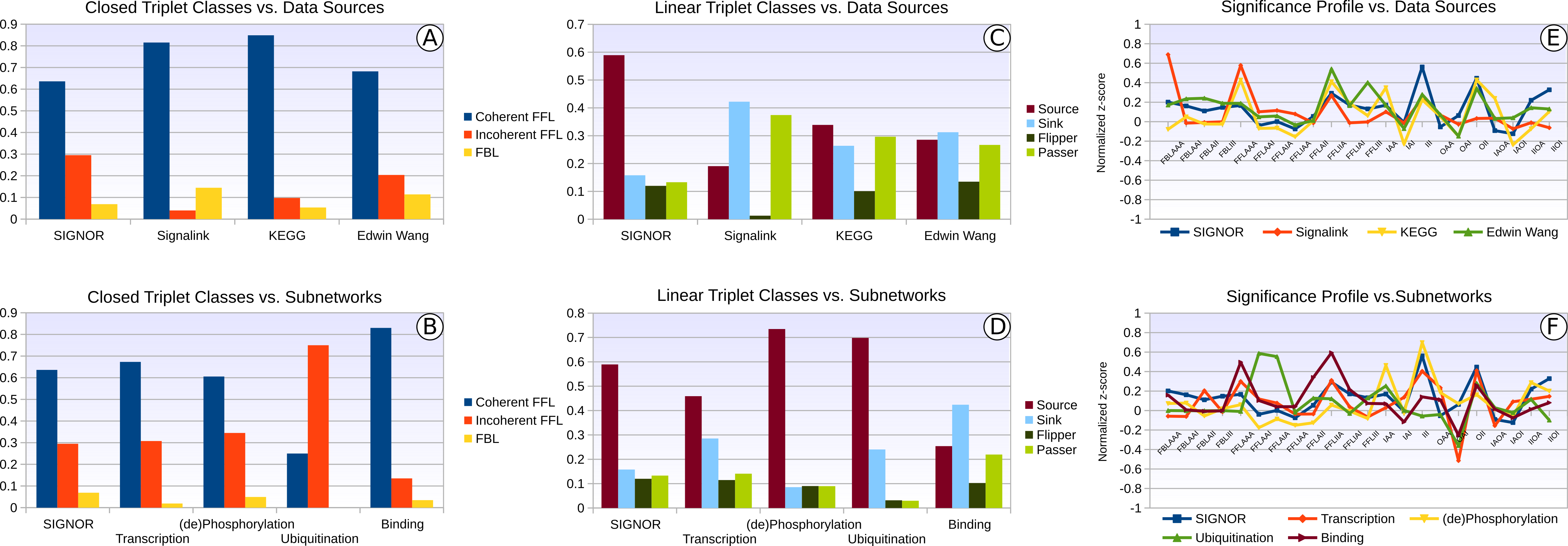}
\caption{{\bf Comparison of motifs abundance in signaling networks and data sources expressed as fractions} Significance profiles for different subnetworks and data sources (E, F). A and B show motif fractions for the major \textit{Closed Triplets} classes. Feed forward and in general more abundant than feedback loops no matter the data source (A). In particular incoherent feed forward are more abundant in ubiquitination subnetwork (B). C and D show motif fractions for the major \textit{Linear Triplets} classes. Flippers are always the least abundant class no matter the database and subnetwork considered (C, D) while, with the exception of Binding, Sources are the most abundant class (D). Significance profiles for different data sources show a similarity among different networks despite curation emphasis (E). Different signaling network are similar but they do exhibit distinctive motifs, suggesting that certain motifs are related to specific functions (F).}
\label{Figure2}
\end{figure}

First we compared the different data sources confirming a similar relative abundance between feed forward and feedback loops, and between Passers and Flippers (Fig~\ref{Figure2} A and C). Alon and co-workers were the first to analyze network motifs in E. coli or S. cerevisiae transcriptional networks. This work was then followed by the same group and other researches \cite{Shen-Orr2002} \cite{Milo2002} \cite{Mangan2003} \cite{Vinayagam2011} \cite{Alon2007a} \cite{Alon2007}. As a first step, we aimed at extending the con-clusions drawn in these reports to a mammalian transcriptional network.

In order to derive a mammalian transcriptional network we used the SIGNOR database since it also annotates the nature of each interaction, i.e. if an interaction is a transcriptional regulation, a phosphorylation etc. In principle, also SignaLink contains information about transcriptional interactions but the positive or negative effect is not annotated, thus preventing the extension of the analysis to this dataset.

Looking at the relative abundance of each motif in the transcriptional networks derived from SIGNOR (Fig~\ref{Figure2}A) we conclude that in high eukaryotes, as is S. cerevisiae, transcriptional networks feed forward loops are more abundant than feedback loops and that incoherent loops are more rare than the coherent ones (Fig~\ref{Figure2}B).

Thanks to the curation richness of the SIGNOR dataset we could also perform similar analyses on subnetworks containing only relationships based on specific molecular mechanisms (transcription, phosphorylation, ubiquitination, binding). These analyses allowed us to generalize conclusions that have already been reported by highlighting that feed forward loops are most abundant in most considered subnetworks with the exception of ubiquitination. This abundance of feed forward loops is not sur-prising and agrees with the principle of minimum energy: it takes more energy to stop a cascade of events than to modulate one. It is interesting to notice how incoherent loops are frequent in ubiquitination.

Finally, to obtain a global comparison of signaling networks, despite their difference in size, we traced significance profiles for the four analyzed datasets (Fig~\ref{Figure2}E) and for the SIGNOR subnetworks (Fig~\ref{Figure2}F). This analysis further confirms the similarity of the different networks and makes it less likely that different curation emphasis may affect our conclusions. In particular, Fig~\ref{Figure2}F shows differences in profiles among subnetworks suggesting once again a relationship between motifs and function.

To our knowledge, no analysis of \textit{Linear Triplets} in biological networks has been reported yet. Through our analysis we could make considerations about the roles played by the nodes involved in such motifs.

In Fig~\ref{Figure2} C and D, we analyze \textit{Linear Triplets}. The most striking observation is the abundance of source motifs in general with the exception of the “binding network”.

\begin{figure}[htp]
\centering
\includegraphics[scale=0.15]{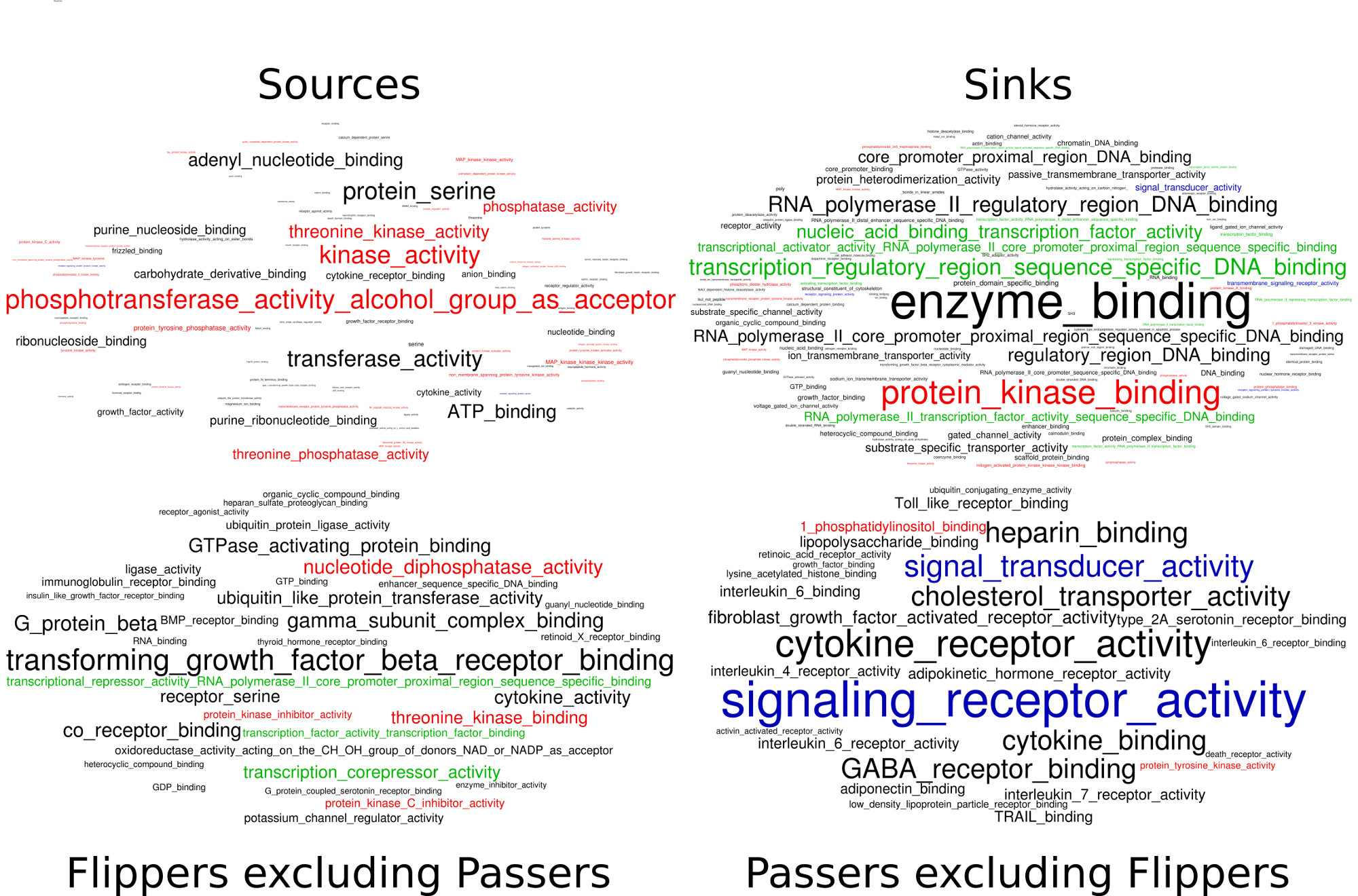}
\caption{{\bf Word clouds of Gene Ontology terms that are annotated to proteins that  are  observed in different classes of \textit{Linear Motifs} (open triangles)}. Word size is proportional to the significance of a term while word color is assigned according to the presence of specific terms. Phosphatases/Kinases terms are in red, transcription related terms are in green, signaling terms in blue and in black all other terms.}
\label{Figure3}
\end{figure}

To ask whether proteins with different functional annotation would preferentially participate  in different motifs we performed a GO \cite{Blake2013} molecular function term enrichment analysis of proteins participating in the formation of the different classes of linear motifs. The results are represented in Fig~\ref{Figure3} as word clouds. 

Nodes that are involved in Source motifs are preferentially annotated with terms related to regulatory phosphorylation events. Nodes involved in Sink motifs are more frequently annotated with transcriptional terms. Nodes which are exclusively involved in Flipper motifs are connected with transcriptional and regulatory events, while nodes which are exclusively involved in Passer motifs are often related to signal transduction mediated by membrane receptors.

Proteins which are more often observed in source motifs are preferentially regulatory proteins, as demonstrated by the presence of terms containing phosphorylation keywords, terms colored in red. On the other hand, proteins, which tend to be in a sink motif are involved in regulation of transcription, terms colored in green. 

As far as Flippers and Passers are concerned, we intersected the two sets and analyzed those nodes that are Flippers but never Passers and vice versa. Nodes that are exclusively Passers tend to be related to receptors and membranes, which are the starting point of signal transduction cascades. On the other hand, Flippers are related to the final steps of signal transduction, transcription regulation.

\subsection*{Using motif profiles as features to build classifiers of protein function}
In order to assess if protein motif profiles underlie molecular function, we applied machine learning to derive a model to infer molecular functions from the motif profile annotated as features of any given node.

By using as node features raw data grouped into six classes (Coherent FFL, Incoherent FFL, FBL, Source, Sinks, Flippers and Passers) and four protein classes derived from UniProt annotation (Phosphatases, Kinases, Transcription Factors, Receptors) we were able to obtain a Random Forest classifier with a Cohen's K of 0.47, which indicates a moderate agreement between predictions and known classes \cite{McHugh2012}.

One interesting considerations about the classifier is the variable importance of the different motif classes. We concluded that source, passer and flipper motifs are relatively more important that sink motifs; result which is compatible with the simple idea that the function of a protein is linked to its outgoing effects and not its ingoing signals (Supplementary Figure 1).
This preliminary machine learning analysis suggests a connection between a combination of network motifs abundance and molecular function. We also built a Random Forest classifier of protein function by using all features without grouping them. Also in this case we obtained a classifier with a Cohen's K of 0.52, which is slightly better than the one obtained with absolute values suggesting that when more data will be available these kind of approaches can be further investigated.

\section*{Conclusion}

While much is known about global characteristics of many biological networks either oriented or not, little is known about the topological properties of signaling networks. Signaling networks are oriented and each interaction has an effect on its target, either positive or negative. Network motifs represent the fine structure of networks which might be linked to the protein function. In this work we analyzed network motifs of three elements, \textit{Linear} and \textit{Closed Triplets}, and we showed that the properties already described for E. coli and S. cerevisiae  transcriptional networks can be extended to other signaling networks. In addition we provide some evidence that the local topology of any specific node is related to its molecular function.

We showed for the first time the role of \textit{Linear Triplets}: Source, Sink, Passer and Flipper. We came to the conclusion that network motif regularities, though the data currently available is scarce and unbalanced, can be used to infer protein function through machine learning approaches highlighting a correlation between topology and functions.

Finally, we release our motifs counter as a stand-alone application and as a Cytoscape app in order to promote further investigations in biological and non-biological networks (Supplementary Martial).


\section*{Competing interests}
  The authors declare that they have no competing interests.

\section*{Acknowledgements}
The authors would like to thank Elisa Micarelli, Daniele Santoni and Paola Bertolazzi for the technical support and coopration, and Livia Perfetto for the insightful biological conversations.

\section*{Funding}
This work has been supported by  the DEPTH grant from the European Research Council (grant agreement 322749) and from a grant from the Italian association from cancer research AIRC(IG 2017, Id. 20322)  to GC.


\bibliographystyle{bmc-mathphys} 


\begin{thebibliography}{13}
\ifx \bisbn   \undefined \def \bisbn  #1{ISBN #1}\fi
\ifx \binits  \undefined \def \binits#1{#1}\fi
\ifx \bauthor  \undefined \def \bauthor#1{#1}\fi
\ifx \batitle  \undefined \def \batitle#1{#1}\fi
\ifx \bjtitle  \undefined \def \bjtitle#1{#1}\fi
\ifx \bvolume  \undefined \def \bvolume#1{\textbf{#1}}\fi
\ifx \byear  \undefined \def \byear#1{#1}\fi
\ifx \bissue  \undefined \def \bissue#1{#1}\fi
\ifx \bfpage  \undefined \def \bfpage#1{#1}\fi
\ifx \blpage  \undefined \def \blpage #1{#1}\fi
\ifx \burl  \undefined \def \burl#1{\textsf{#1}}\fi
\ifx \doiurl  \undefined \def \doiurl#1{\textsf{#1}}\fi
\ifx \betal  \undefined \def \betal{\textit{et al.}}\fi
\ifx \binstitute  \undefined \def \binstitute#1{#1}\fi
\ifx \binstitutionaled  \undefined \def \binstitutionaled#1{#1}\fi
\ifx \bctitle  \undefined \def \bctitle#1{#1}\fi
\ifx \beditor  \undefined \def \beditor#1{#1}\fi
\ifx \bpublisher  \undefined \def \bpublisher#1{#1}\fi
\ifx \bbtitle  \undefined \def \bbtitle#1{#1}\fi
\ifx \bedition  \undefined \def \bedition#1{#1}\fi
\ifx \bseriesno  \undefined \def \bseriesno#1{#1}\fi
\ifx \blocation  \undefined \def \blocation#1{#1}\fi
\ifx \bsertitle  \undefined \def \bsertitle#1{#1}\fi
\ifx \bsnm \undefined \def \bsnm#1{#1}\fi
\ifx \bsuffix \undefined \def \bsuffix#1{#1}\fi
\ifx \bparticle \undefined \def \bparticle#1{#1}\fi
\ifx \barticle \undefined \def \barticle#1{#1}\fi
\ifx \bconfdate \undefined \def \bconfdate #1{#1}\fi
\ifx \botherref \undefined \def \botherref #1{#1}\fi
\ifx \url \undefined \def \url#1{\textsf{#1}}\fi
\ifx \bchapter \undefined \def \bchapter#1{#1}\fi
\ifx \bbook \undefined \def \bbook#1{#1}\fi
\ifx \bcomment \undefined \def \bcomment#1{#1}\fi
\ifx \oauthor \undefined \def \oauthor#1{#1}\fi
\ifx \citeauthoryear \undefined \def \citeauthoryear#1{#1}\fi
\ifx \endbibitem  \undefined \def \endbibitem {}\fi
\ifx \bconflocation  \undefined \def \bconflocation#1{#1}\fi
\ifx \arxivurl  \undefined \def \arxivurl#1{\textsf{#1}}\fi
\csname PreBibitemsHook\endcsname

\bibitem{koon}
\begin{barticle}
\bauthor{\bsnm{Koonin}, \binits{E.V.}},
\bauthor{\bsnm{Altschul}, \binits{S.F.}},
\bauthor{\bsnm{Bork}, \binits{P.}}:
\batitle{Brca1 protein products: functional motifs}.
\bjtitle{Nat Genet}
\bvolume{13},
\bfpage{266}--\blpage{267}
(\byear{1996})
\end{barticle}
\endbibitem

\bibitem{khar}
\begin{botherref}
\oauthor{\bsnm{Kharitonov}, \binits{S.A.}},
\oauthor{\bsnm{Barnes}, \binits{P.J.}}:
Clinical Aspects of Exhaled Nitric Oxide.
in press
\end{botherref}
\endbibitem

\bibitem{zvai}
\begin{barticle}
\bauthor{\bsnm{Zvaifler}, \binits{N.J.}},
\bauthor{\bsnm{Burger}, \binits{J.A.}},
\bauthor{\bsnm{Marinova-Mutafchieva}, \binits{L.}},
\bauthor{\bsnm{Taylor}, \binits{P.}},
\bauthor{\bsnm{Maini}, \binits{R.N.}}:
\batitle{Mesenchymal cells, stromal derived factor-1 and rheumatoid arthritis
  [abstract]}.
\bjtitle{Arthritis Rheum}
\bvolume{42},
\bfpage{250}
(\byear{1999})
\end{barticle}
\endbibitem

\bibitem{xjon}
\begin{bchapter}
\bauthor{\bsnm{Jones}, \binits{X.}}:
\bctitle{Zeolites and synthetic mechanisms}.
In: \beditor{\bsnm{Smith}, \binits{Y.}} (ed.)
\bbtitle{Proceedings of the First National Conference on Porous Sieves: 27-30
  June 1996; Baltimore},
pp. \bfpage{16}--\blpage{27}
(\byear{1996}).
\bcomment{Stoneham: Butterworth-Heinemann}
\end{bchapter}
\endbibitem

\bibitem{marg}
\begin{bbook}
\bauthor{\bsnm{Margulis}, \binits{L.}}:
\bbtitle{Origin of Eukaryotic Cells}.
\bpublisher{Yale University Press},
\blocation{New Haven}
(\byear{1970})
\end{bbook}
\endbibitem

\bibitem{oreg}
\begin{barticle}
\bauthor{\bsnm{Orengo}, \binits{C.A.}},
\bauthor{\bsnm{Bray}, \binits{J.E.}},
\bauthor{\bsnm{Hubbard}, \binits{T.}},
\bauthor{\bsnm{LoConte}, \binits{L.}},
\bauthor{\bsnm{Sillitoe}, \binits{I.}}:
\batitle{Analysis and assessment of ab initio three-dimensional prediction,
  secondary structure, and contacts prediction}.
\bjtitle{Proteins}
\bvolume{Suppl 3},
\bfpage{149}--\blpage{170}
(\byear{1999})
\end{barticle}
\endbibitem

\bibitem{schn}
\begin{bchapter}
\bauthor{\bsnm{Schnepf}, \binits{E.}}:
\bctitle{From prey via endosymbiont to plastids: comparative studies in
  dinoflagellates}.
In: \beditor{\bsnm{Lewin}, \binits{R.A.}} (ed.)
\bbtitle{Origins of Plastids}
vol. \bseriesno{2},
\bedition{2nd} edn.,
pp. \bfpage{53}--\blpage{76}.
\bpublisher{Chapman and Hall},
\blocation{New York}
(\byear{1993})
\end{bchapter}
\endbibitem

\bibitem{pond}
\begin{botherref}
Innovative Oncology
\end{botherref}
\endbibitem

\bibitem{smith}
\begin{bbook}
\beditor{\bsnm{Smith}, \binits{Y.}} (ed.):
\bbtitle{Proceedings of the First National Conference on Porous Sieves: 27-30
  June 1996; Baltimore}.
\bpublisher{Butterworth-Heinemann},
\blocation{Stoneham}
(\byear{1996})
\end{bbook}
\endbibitem

\bibitem{hunn}
\begin{bchapter}
\bauthor{\bsnm{Hunninghake}, \binits{G.W.}},
\bauthor{\bsnm{Gadek}, \binits{J.E.}}:
\bctitle{The alveloar macrophage}.
In: \beditor{\bsnm{Harris}, \binits{T.J.R.}} (ed.)
\bbtitle{Cultured Human Cells and Tissues},
pp. \bfpage{54}--\blpage{56}.
\bpublisher{Academic Press},
\blocation{New York}
(\byear{1995}).
\bcomment{Stoner G (Series Editor): Methods and Perspectives in Cell Biology,
  vol 1}
\end{bchapter}
\endbibitem

\bibitem{advi}
\begin{botherref}
Advisory Committee on Genetic Modification:
Annual Report.
London
(1999).
Advisory Committee on Genetic Modification
\end{botherref}
\endbibitem

\bibitem{koha}
\begin{botherref}
\oauthor{\bsnm{Kohavi}, \binits{R.}}:
Wrappers for performance enhancement and obvious decision graphs.
PhD thesis,
Stanford University, Computer Science Department
(1995)
\end{botherref}
\endbibitem

\bibitem{mouse}
\begin{botherref}
The Mouse Tumor Biology Database.
\url{http://tumor.informatics.jax.org/cancer\_links.html}
\end{botherref}
\endbibitem

\end{thebibliography}

\newcommand{\BMCxmlcomment}[1]{}

\BMCxmlcomment{

<refgrp>

<bibl id="B1">
  <title><p>BRCA1 protein products: functional motifs</p></title>
  <aug>
    <au><snm>Koonin</snm><fnm>E V</fnm></au>
    <au><snm>Altschul</snm><fnm>S F</fnm></au>
    <au><snm>Bork</snm><fnm>P</fnm></au>
  </aug>
  <source>Nat Genet</source>
  <pubdate>1996</pubdate>
  <volume>13</volume>
  <fpage>266</fpage>
  <lpage>267</lpage>
</bibl>

<bibl id="B2">
  <title><p>Clinical aspects of exhaled nitric oxide</p></title>
  <aug>
    <au><snm>Kharitonov</snm><fnm>S A</fnm></au>
    <au><snm>Barnes</snm><fnm>P J</fnm></au>
  </aug>
  <source>Eur Respir J</source>
  <inpress />
</bibl>

<bibl id="B3">
  <title><p>Mesenchymal cells, stromal derived factor-1 and rheumatoid
  arthritis [abstract]</p></title>
  <aug>
    <au><snm>Zvaifler</snm><fnm>N J</fnm></au>
    <au><snm>Burger</snm><fnm>J A</fnm></au>
    <au><snm>Marinova Mutafchieva</snm><fnm>L</fnm></au>
    <au><snm>Taylor</snm><fnm>P</fnm></au>
    <au><snm>Maini</snm><fnm>R N</fnm></au>
  </aug>
  <source>Arthritis Rheum</source>
  <pubdate>1999</pubdate>
  <volume>42</volume>
  <fpage>s250</fpage>
</bibl>

<bibl id="B4">
  <title><p>Zeolites and synthetic mechanisms</p></title>
  <aug>
    <au><snm>Jones</snm><fnm>X</fnm></au>
  </aug>
  <source>Proceedings of the First National Conference on Porous Sieves: 27-30
  June 1996; Baltimore</source>
  <editor>Y Smith</editor>
  <pubdate>1996</pubdate>
  <fpage>16</fpage>
  <lpage>27</lpage>
</bibl>

<bibl id="B5">
  <title><p>Origin of Eukaryotic Cells</p></title>
  <aug>
    <au><snm>Margulis</snm><fnm>L</fnm></au>
  </aug>
  <publisher>New Haven: Yale University Press</publisher>
  <pubdate>1970</pubdate>
</bibl>

<bibl id="B6">
  <title><p>Analysis and assessment of ab initio three-dimensional prediction,
  secondary structure, and contacts prediction</p></title>
  <aug>
    <au><snm>Orengo</snm><fnm>C A</fnm></au>
    <au><snm>Bray</snm><fnm>J E</fnm></au>
    <au><snm>Hubbard</snm><fnm>T</fnm></au>
    <au><snm>LoConte</snm><fnm>L</fnm></au>
    <au><snm>Sillitoe</snm><fnm>I</fnm></au>
  </aug>
  <source>Proteins</source>
  <pubdate>1999</pubdate>
  <volume>Suppl 3</volume>
  <fpage>149</fpage>
  <lpage>170</lpage>
</bibl>

<bibl id="B7">
  <title><p>From prey via endosymbiont to plastids: comparative studies in
  dinoflagellates</p></title>
  <aug>
    <au><snm>Schnepf</snm><fnm>E</fnm></au>
  </aug>
  <source>Origins of Plastids</source>
  <publisher>New York: Chapman and Hall</publisher>
  <editor>R A Lewin</editor>
  <edition>2</edition>
  <pubdate>1993</pubdate>
  <volume>2</volume>
  <fpage>53</fpage>
  <lpage>76</lpage>
</bibl>

<bibl id="B8">
  <title><p>Innovative oncology</p></title>
  <source>Breast Cancer Res</source>
  <editor>B Ponder and S Johnston and L Chodosh</editor>
  <pubdate>1998</pubdate>
  <volume>10</volume>
  <fpage>1</fpage>
  <lpage>72</lpage>
</bibl>

<bibl id="B9">
  <title><p>Proceedings of the First National Conference on Porous Sieves:
  27-30 June 1996; Baltimore</p></title>
  <publisher>Stoneham: Butterworth-Heinemann</publisher>
  <editor>Y Smith</editor>
  <pubdate>1996</pubdate>
</bibl>

<bibl id="B10">
  <title><p>The alveloar macrophage</p></title>
  <aug>
    <au><snm>Hunninghake</snm><fnm>G W</fnm></au>
    <au><snm>Gadek</snm><fnm>J E</fnm></au>
  </aug>
  <source>Cultured Human Cells and Tissues</source>
  <publisher>New York: Academic Press</publisher>
  <editor>T J R Harris</editor>
  <pubdate>1995</pubdate>
  <fpage>54</fpage>
  <lpage>56</lpage>
  <note>Stoner G (Series Editor): Methods and Perspectives in Cell Biology, vol
  1</note>
</bibl>

<bibl id="B11">
  <title><p>Annual Report</p></title>
  <aug><au><cnm>Advisory Committee on Genetic Modification</cnm></au></aug>
  <publisher>London</publisher>
  <pubdate>1999</pubdate>
</bibl>

<bibl id="B12">
  <title><p>Wrappers for performance enhancement and obvious decision
  graphs</p></title>
  <aug>
    <au><snm>Kohavi</snm><fnm>R</fnm></au>
  </aug>
  <source>PhD thesis</source>
  <publisher>Stanford University, Computer Science Department</publisher>
  <pubdate>1995</pubdate>
</bibl>

<bibl id="B13">
  <title><p>The Mouse Tumor Biology Database</p></title>
  <url>http://tumor.informatics.jax.org/cancer\_links.html</url>
</bibl>

</refgrp>
} 


\begin{thebibliography}{10}

\bibitem{Girvan2002}
Girvan M, Newman MEJ, Girvan M, Newman MEJ, Newman MEJ.
\newblock {Community structure in social and biological networks.}
\newblock Proceedings of the National Academy of Sciences of the United States
  of America. 2002;99(12):7821--6.
\newblock doi:{10.1073/pnas.122653799}.

\bibitem{Maslov2002}
Maslov S, Sneppen K.
\newblock {Specificity and Stability in Topology of Protein Networks}.
\newblock Science. 2002;296(5569):910--913.
\newblock doi:{10.1126/science.1065103}.

\bibitem{Calderone2016}
Calderone A, Formenti M, Aprea F, Papa M, Alberghina L, Colangelo AM, et~al.
\newblock {Comparing Alzheimer's and Parkinson's diseases networks using graph
  communities structure}.
\newblock BMC Systems Biology. 2016;10(1):25.
\newblock doi:{10.1186/s12918-016-0270-7}.

\bibitem{Schwikowski2000}
Schwikowski B, Uetz P, Fields S.
\newblock {A network of protein-protein interactions in yeast.}
\newblock Nature biotechnology. 2000;18(12):1257--61.
\newblock doi:{10.1038/82360}.

\bibitem{Hartwell}
Hartwell LH, Hopfield JJ, Leibler S, Murray AW.
\newblock {From molecular to modular cell biology.}
\newblock Nature. 1999;402(6761 Suppl):C47--52.
\newblock doi:{10.1038/35011540}.

\bibitem{Bader2003}
Bader GD, Hogue CWV.
\newblock {An automated method for finding molecular complexes in large protein
  interaction networks.}
\newblock BMC bioinformatics. 2003;4:2.

\bibitem{Scott2006}
Scott J, Ideker T, Karp RM, Sharan R.
\newblock {Efficient algorithms for detecting signaling pathways in protein
  interaction networks.}
\newblock Journal of computational biology : a journal of computational
  molecular cell biology. 2006;13(2):133--44.
\newblock doi:{10.1089/cmb.2006.13.133}.

\bibitem{Vazquez2004}
V{\'{a}}zquez A, Dobrin R, Sergi D, Eckmann JP, Oltvai ZN, Barab{\'{a}}si AL.
\newblock {The topological relationship between the large-scale attributes and
  local interaction patterns of complex networks.}
\newblock Proceedings of the National Academy of Sciences of the United States
  of America. 2004;101(52):17940--5.
\newblock doi:{10.1073/pnas.0406024101}.

\bibitem{Shen-Orr2002}
Shen-Orr SS, Milo R, Mangan S, Alon U.
\newblock {Network motifs in the transcriptional regulation network of
  Escherichia coli.}
\newblock Nature genetics. 2002;31(1):64--8.
\newblock doi:{10.1038/ng881}.

\bibitem{Milo2002}
Milo R, Shen-Orr S, Itzkovitz S, Kashtan N, Chklovskii D, Alon U.
\newblock {Network Motifs: Simple Building Blocks of Complex Networks}.
\newblock Science. 2002;298(5594):824--827.
\newblock doi:{10.1126/science.298.5594.824}.

\bibitem{McAdams1995}
McAdams H, Shapiro L.
\newblock {Circuit simulation of genetic networks}.
\newblock Science. 1995;269(5224):650--656.
\newblock doi:{10.1126/science.7624793}.

\bibitem{Mangan2003}
Mangan S, Alon U.
\newblock {Structure and function of the feed-forward loop network motif.}
\newblock Proceedings of the National Academy of Sciences of the United States
  of America. 2003;100(21):11980--11985.
\newblock doi:{10.1073/pnas.2133841100}.

\bibitem{Vinayagam2014}
Vinayagam A, Zirin J, Roesel C, Hu Y, Yilmazel B, Samsonova AA, et~al.
\newblock {Integrating protein-protein interaction networks with phenotypes
  reveals signs of interactions.}
\newblock Nature methods. 2014;11(1):94--9.
\newblock doi:{10.1038/nmeth.2733}.

\bibitem{Orchard2014}
Orchard S, Ammari M, Aranda B, Breuza L, Briganti L, Broackes-Carter F, et~al.
\newblock {The MIntAct project--IntAct as a common curation platform for 11
  molecular interaction databases.}
\newblock Nucleic acids research. 2014;42(1):D358--63.
\newblock doi:{10.1093/nar/gkt1115}.

\bibitem{Calderone2013}
Calderone A, Castagnoli L, Cesareni G.
\newblock {Mentha: a Resource for Browsing Integrated Protein-Interaction
  Networks.}
\newblock Nature methods. 2013;10(8):690.
\newblock doi:{10.1038/nmeth.2561}.

\bibitem{Perfetto2015}
Perfetto L, Briganti L, Calderone A, Perpetuini AC, Iannuccelli M, Langone F,
  et~al.
\newblock {SIGNOR: a database of causal relationships between biological
  entities.}
\newblock Nucleic acids research. 2015;doi:{10.1093/nar/gkv1048}.

\bibitem{Kanehisa2000}
Kanehisa M, Goto S.
\newblock {KEGG: Kyoto Encyclopedia of Genes and Genomes}.
\newblock Nucleic Acids Research. 2000;28(1):27--30.

\bibitem{Fazekas2013}
Fazekas D, Koltai M, T{\"{u}}rei D, M{\'{o}}dos D, P{\'{a}}lfy M, D{\'{u}}l Z,
  et~al.
\newblock {SignaLink 2 - a signaling pathway resource with multi-layered
  regulatory networks.}
\newblock BMC systems biology. 2013;7(1):7.
\newblock doi:{10.1186/1752-0509-7-7}.

\bibitem{Zaman2013}
Zaman N, Li L, Jaramillo M, Sun Z, Tibiche C, Banville M, et~al.
\newblock {Signaling Network Assessment of Mutations and Copy Number Variations
  Predict Breast Cancer Subtype-Specific Drug Targets}.
\newblock Cell Reports. 2013;5(1):216--223.
\newblock doi:{10.1016/j.celrep.2013.08.028}.

\bibitem{OMadadhain2005}
O'Madadhain J, Fisher D, Padhraic S, Boey YB, White S, {Joshua O'Madadhain}
  SWPSYbB Danyel~Fisher, et~al.
\newblock {Analysis and visualization of network data using JUNG.}
\newblock Journal of Statistical Software. 2005;VV:1--35.

\bibitem{Milo2004}
Milo R, Itzkovitz S, Kashtan N, Levitt R, Shen-Orr S, Ayzenshtat I, et~al.
\newblock {Superfamilies of Evolved and Designed Networks}.
\newblock Science. 2004;303(5663):1538--1542.
\newblock doi:{10.1126/science.1089167}.

\bibitem{Wong2012}
Wong E, Baur B, Quader S, Huang CH.
\newblock {Biological network motif detection: Principles and practice}.
\newblock Briefings in Bioinformatics. 2012;13(2):202--215.
\newblock doi:{10.1093/bib/bbr033}.

\bibitem{Kuhn2008}
Kuhn M.
\newblock {caret Package}.
\newblock Journal Of Statistical Software. 2008;.

\bibitem{Akaike1998}
Akaike H.
\newblock {Information Theory and an Extension of the Maximum Likelihood
  Principle}.
\newblock Springer New York; 1998. p. 199--213.
\newblock Available from:
  \url{http://link.springer.com/10.1007/978-1-4612-1694-0{\_}15}.

\bibitem{Vinayagam2011}
Vinayagam A, Stelzl U, Foulle R, Plassmann S, Zenkner M, Timm J, et~al.
\newblock {A directed protein interaction network for investigating
  intracellular signal transduction.}
\newblock Science signaling. 2011;4(189):rs8.

\bibitem{Alon2007a}
Alon U.
\newblock {Network motifs: theory and experimental approaches.}
\newblock Nature reviews Genetics. 2007;8(6):450--61.
\newblock doi:{10.1038/nrg2102}.

\bibitem{Alon2007}
Alon U.
\newblock {An Introduction to Systems Biology: Design Principles of Biological
  Circuits}. vol.~10 of Chapman {\&} Hall/CRC mathematical and computational
  biology series.
\newblock Chapman{\{}{\&}{\}}Hall/CRC; 2007.
\newblock Available from:
  \url{http://www.loc.gov/catdir/enhancements/fy0654/2005056902-d.html}.

\bibitem{Blake2013}
Blake JA, Dolan M, Drabkin H, Hill DP, Li N, Sitnikov D, et~al.
\newblock {Gene Ontology annotations and resources.}
\newblock Nucleic acids research. 2013;41(Database issue):D530--5.
\newblock doi:{10.1093/nar/gks1050}.

\bibitem{McHugh2012}
McHugh ML.
\newblock {Interrater reliability: the kappa statistic.}
\newblock Biochemia medica. 2012;22(3):276--82.

\end{thebibliography}




\end{document}